\documentclass[runningheads]{llncs}
\usepackage{graphicx}
\DeclareUnicodeCharacter{2212}{-}
\begin{document}
\title{Hyperparameter Optimization for COVID-19 Chest X-Ray Classification}
%
%
\author{Ibraheem Hamdi \and
Muhammad Ridzuan \and
Mohammad Yaqub}
\authorrunning{Ibraheem Hamdi et al.}
%
\institute{Mohamed bin Zayed University of Artificial Intelligence, Abu Dhabi, U.A.E}
\maketitle              
\begin{abstract}
Despite the introduction of vaccines, Coronavirus disease (COVID-19) remains a worldwide dilemma, continuously developing new variants such as Delta and the recent Omicron. The current standard for testing is through polymerase chain reaction (PCR). However, PCRs can be expensive, slow, and/or inaccessible to many people. X-rays on the other hand have been readily used since the early 20th century and are relatively cheaper, quicker to obtain, and typically covered by health insurance. With a careful selection of model, hyperparameters, and augmentations, we show that it is possible to develop models with 83\% accuracy in binary classification and 64\% in multi-class for detecting COVID-19 infections from chest x-rays.

\end{abstract}
\section{Introduction}
COVID-19 is an infectious disease that causes mild to moderate respiratory illness, and in severe cases death. Over 2.5 million new cases were reported in the last 24 hours~\cite{ref_url1}, and 60\% of the US population are expected to be infected by March of 2022~\cite{ref_url2}. 

PCR, a molecular test that can identify the disease in genetic material from collected swabs, is commonly used to detect the disease~\cite{ref_url3}. However, the test can cost over 400 US dollars~\cite{ref_url4}, and results are taking longer and longer due to the rise in cases~\cite{ref_url5,ref_url6,ref_url7}. Therefore, it is becoming increasingly important to develop a relatively cheaper, faster, and more accessible screening procedure.

X-rays were discovered in the late 19th century and have been used in the medical field ever since~\cite{ref_url8}. Chest x-rays can be obtained within 15 minutes, including positioning time~\cite{ref_url9}, making them rather quick to produce for a rapid diagnosis. To create a model that can recognize and classify COVID-19 infections from chest x-rays would be immensely beneficial, especially at times like these when early quarantine is critical to limit the spread the spread of the virus.

For learning purposes, we made use of vanilla DenseNet-121~\cite{ref_paper1} architecture as well as TorchXRayVision~\cite{ref_url10} library’s variations that are pre-trained on x-ray data from different data combinations. We investigated the effect of model, hyperparameter, and augmentation choice on the accuracy of classification. In the beginning, the program was written as a Jupyter Notebook using only vanilla DenseNet-121 and without any augmentations. Each experiment required manual modification to the code in multiple locations. With the introduction of TorchXrayVision's models and more augmentations, making a mistake while modifying the code became easier and tracebacks errors started happening. The solution to that was adding a new section in the beginning of the code that asked the user what augmentations, hyperparameters, and models are required. At that time, the only way to stop the code without reaching the set maximum number of epochs was to physically interrupt the program. Furthermore, model checkpoints were being created at every epoch, slowing down the training process and using up too much storage space. To address that, a user-defined validation accuracy was set to start saving checkpoints and another to stop the program.

As the project grew, it became more obvious that a better system was needed, one that is more automated. Preferably, it would have features like early-stopping, to save the best model and retain GPU time, and online logging, to enable remote monitoring. Luckily, a presentation by one of the students at the university introduced us to a potential solution. Hydra-Lightning-Template~\cite{ref_url11} became the core of this program. It enables easy experimentation, hyperparameter search with Optuna~\cite{ref_url12}, and online logging through Weights \& Biases (Wandb)~\cite{ref_url13}. It is built on Pytorch Lightning, which relies on ‘.yaml’ configuration files to retrieve experiment settings such as hyperparameters and model. This immensely simplified experimentation by minimizing the amount of boilerplate code, allowing us to produce different models without the need to manually change the code.

Optuna is an automatic hyperparameter optimization framework that is integrated into Pytorch Lightning. It has a simple and intuitive interface that is platform agnostic and lightweight with minimal infrastructure dependency, making it extremely versatile. Features include pruning to terminate unpromising trials and speed up optimization, easy parallelization to enable scalability across multiple machines, quick visualization through a built-in dashboard, as well as an abundance of optimization samplers~\cite{ref_url18}. This makes Optuna more suitable for our experimentation rather than other methods like GPyOpt, SageMaker, or GoogleVizier, that don't have all of these advantages and features together~\cite{ref_url17}.

\section{Dataset}
The Society for Imaging Informatics in Medicine (SIIM) partnered with the Foundation for the Promotion of Health and Biomedical Research of Valencia Region (FISABIO), Medical Imaging Databank of Valencia Region (BIMCV) and the Radiological Society of North America (RSNA) to create a Kaggle competition with the aim to advance the identification of COVID-19 Pneumonia from chest x-rays. 

The dataset provided is composed of over 6000 scans in DICOM format belonging to four different classes: 1,676 ‘Negative for Pneumonia’, 2,855 ‘Typical Appearance’, 1,049 ‘Indeterminate Appearance’, and 474 ‘Atypical Appearance’~\cite{ref_url14}. To ensure data balance for binary experiments, 50\% of the data was composed of ‘Negative for Pneumonia’, while the other half was a random mix of the other three classes.

To avoid confusing the model with lateral x-rays, 230 data folders that had more than one DICOM file were excluded. Furthermore, the dataset was stratified according to the assigned labels to ensure a balanced distribution of classes. In our experiments, we used a 70:20:10 ratio to split the data into training, validation, and testing datasets.

\section{Methods}
Optuna offers flexibility through a wide range of samplers to perform the optimization. These include Tree-structured Parzen Estimator (TPA), Gaussian Processes (GP), Covariance Matrix Adaptation Evolution Strategy (CMA-ES), and Random and Grid Search~\cite{ref_url19}. TPA is the default and recommended sampler by Optuna especially for experiments with less than 1000 trials and hence was used. 

Within the Hydra-Lightning-Template, a configuration file can be created to define the optimization settings. For instance, number of trials, optimization metric, sampler or optimization technique, hyperparameters, and range of search are specified inside the '.yaml' file. Optuna then uses the sampler to maximize (or minimize) the optimization metric, validation accuracy in our case, by trying different hyperparameter values within the search space. Based on the learning curve, Optuna may choose to prune or early-stop the experiment to save time and try another set of hyperparameters.

\section{Setup}
\subsection{Image Pre-processing}
Pydicom~\cite{ref_url15} library was used to read the x-ray images provided in DICOM format. The function ‘pixel\_array’ is used to convert the images to a 2-dimensional NumPy array. However, Vanilla DenseNet-121 expects 3-channel input, while TorchXRayVision models expect a grayscale image of size 224 x 224. Therefore, we had to design our Data Module to handle files differently according to the model specified for the experiment.

According to the configuration of each experiment, geometrical transformations such as  scaling, shear, rotation, translation, as well as horizontal and vertical flipping were applied to investigate the effects of augmentations on the accuracy of models. Scaling, shear, rotation, and translation were chosen because they naturally exist in x-ray imaging as a result of positioning error, while horizontal and vertical flipping can happen as a result of input or scanning error. These transformations were chosen as they make a good representation of what is seen in the real world. Color transformations however would not help since x-rays are always produced in grayscale.

\subsection{Models}
DenseNet-121 model was chosen for this task due to its reported success with chest x-rays in literature~\cite{ref_paper2}. In addition, there are nine DenseNet-121 TorchXRayVision models that are pretrained on different x-ray datasets such as CheXpert, NIH, and PadChest~\cite{ref_url10}. We encountered an issue with downloading the weights for the ‘JF Healthcare’ model so it was discarded.

Each of its pretrained models has 18 outputs corresponding to the following pathologies:

\begin{multicols}{3}
\begin{itemize}
    \item Atelectasis
    \item Consolidation 
    \item Infiltration
    \item Pneumothorax
    \item Edema
    \item Emphysema
    \item Fibrosis
    \item Effusion
    \item Pneumonia
    \item Pleural\_Thickening
    \item Cardiomegaly
    \item Nodule
    \item Mass
    \item Hernia
    \item Lung Lesion
    \item Fracture
    \item Lung Opacity
    \item Enlarged Cardiomediastinum
\end{itemize}
\end{multicols}

When calling TorchXRayVision models, the number of classes is specified to 2 or 4 according to the experiment. However, vanilla DenseNet-121 doesn’t have such a feature, therefore the classifier was changed to ‘nn.Linear(num\_features, num\_classes)’ where num\_features is in\_features from the original classifier and num\_classes is set to 2 or 4 depending on whether the experiment is binary or multi-class.

\subsection{Search Space}
To keep the architecture consistent across all experiments, the ResNet model from TorchXRayVision was not used. Furthermore, by using Optuna for hyperparameter optimization, augmentations, batch size, dropout rate, models, and learning rate were experimented with. The search limit was set to 0.0001-0.001 for learning rate, 0\%-20\% for dropout rate, -15\%-15\% rotation, -30\%-30\% scale, -30\%-30\% shear, and 0\%-100\% translation. Horizontal and vertical flipping and batch sizes of 8, 16, 32, 64 and 128 were also studied.

\section{Results and Discussion}
Adam optimizer and cross-entropy loss are used for all of our experiments.
\subsection{Binary Classification}
To begin, the models were trained using a learning rate of 0.0003, and batch size of 32. TorchXRayVision’s “ALL” model has been trained on all different datasets and achieved the highest baseline accuracy of 81.61\% (see Table.~\ref{tab1}). Therefore, “ALL” model was used for our binary experiments.

\begin{table}
\caption{Baseline validation accuracy - binary.}\label{tab1}
\begin{center}
\begin{tabular}{|l|r|}
\hline
{\bfseries Model} & {\bfseries Validation Accuracy (\%)} \\
\hline
DenseNet-121 & 72.14\\
\hline
{\bfseries ALL} & {\bfseries 81.61}\\
\hline
CheXpert & 72.14\\
\hline
MIMIC\_CH & 75.63\\
\hline
MIMIC\_NB & 77.80\\
\hline
NIH & 78.45\\
\hline
PadChest & 79.43\\
\hline
RSNA & 50.38\\
\hline
\end{tabular}
\end{center}
\end{table}

We observed that the maximum validation accuracy peaked using 64 batches, and started decreasing on either side (see Table.~\ref{tab2}). This is due to the fact that the ability of a model to generalize degrades beyond a certain batch size~\cite{ref_paper6}. With validation accuracy of 82.81\%, the decision was made to use batch size of 64 from that point for our binary experiments.

\begin{table}
\caption{Batch size vs validation accuracy - binary.}\label{tab2}
\begin{center}
\begin{tabular}{|l|r|}
\hline
{\bfseries Batch Size} & {\bfseries Validation Accuracy (\%)}\\
\hline
8 & 79.43\\
\hline
16 & 80.52\\
\hline
32 & 80.63\\
\hline
{\bfseries 64} & {\bfseries 82.81}\\
\hline
128 & 81.61 \\
\hline
\end{tabular}
\end{center}
\end{table}

Instead of assuming 0.0003, we started searching for the optimal learning rate using “ALL” model and batch size of 64. Validation accuracy peaked at 1.737 (see Table.~\ref{tab3}) and dropped above or below that. With too low of a learning rate, the model will take much longer to reach the optimal solution and might even underfit the data. On the other hand, having a learning rate that is very high means the model will be unable to converge as the steps are too large~\cite{ref_paper7}. The highest validation accuracy of 82.48\% was achieved with a learning rate of 0.0001737, and we estimate that the optimal learning rate is between 0.00016 and 0.00018.

\begin{table}
\caption{Learning rate vs validation accuracy - binary.}\label{tab3}
\begin{center}
\begin{tabular}{|l|r|}
\hline
{\bfseries Learning Rate (e-4)} & {\bfseries Validation Accuracy (\%)}\\
\hline
0.484 & 77.15\\
\hline
0.894 & 80.47\\
\hline
1.273 & 80.41\\
\hline
{\bfseries 1.737} & {\bfseries 82.84}\\
\hline
1.921 & 82.26\\
\hline
2.916 & 82.15\\
\hline
5.913 & 80.41\\
\hline
8.644 & 78.78\\
\hline
\end{tabular}
\end{center}
\end{table}

Dropout is a technique used to boost accuracy and prevent models from overfitting by randomly dropping units from layers during training~\cite{ref_paper4}. This way, a neural network is forced to learn with incomplete information, enabling it to generalize better. Using “ALL” model, batch size of 64, and learning rate of 0.0001737, the best model achieved a validation accuracy of 80.3\% using 13.06\% dropout rate (see Table.~\ref{tab4}). Contrary to previous intuition, using dropout did not provide an improvement in the validation accuracy. It is possible that the network required more time to learn with dropout, and that a better accuracy could've been achieved with longer training.

\begin{table}
\caption{Dropout rate vs validation accuracy - binary.}\label{tab4}
\begin{center}
\begin{tabular}{|l|r|}
\hline
{\bfseries Dropout Rate (\%)} & {\bfseries Validation Accuracy (\%)}\\
\hline
3.678 &  79.98\\
\hline
6.328 & 79.98\\
\hline
11.91 & 80.20\\
\hline
{\bfseries 13.06} & {\bfseries 80.30}\\
\hline
18.59 & 79.76\\
\hline
\end{tabular}
\end{center}
\end{table}

Due to the decrease in accuracy using dropout, we decided to proceed without any to experiment with augmentations, using “ALL” model, batch size of 64, and learning rate of 0.0001737. The model achieved its highest validation accuracy of 80.85\% using scale of +/- 24.12\%, shear of +/-18.09\%, and translation of 64.25\% without any horizontal or vertical flipping (see Table.~\ref{tab5}).

\begin{table}
\caption{Augmentations vs validation accuracy - binary.}\label{tab5}
\begin{center}
\begin{tabular}{|l|l|l|l|l|l|l|l|}
\hline
{\bfseries Horizontal Flip} & Off & {\bfseries On} & Off & Off & On & Off & On\\
\hline
{\bfseries Vertical Flip} & Off & {\bfseries Off} & Off & On & Off & Off & Off\\
\hline
{\bfseries Rotation (+/- o)} & 14 & {\bfseries 13} & 0 & 0 & 14 & 14 & 13\\
\hline
{\bfseries Scale (+/- \%)} & 13.3 & {\bfseries 9.97} & 24.12 & 17.46 & 5.94 & 26.60 & 19.94\\
\hline
{\bfseries Shear (+/- \%)} & 23.91 &  {\bfseries 20.18} & 18.09 & 20.55 & 9.48 & 28.69 & 24.22\\
\hline
{\bfseries Translation (+/- \%)} & 46.19 & {\bfseries 62.71} & 64.25 & 89.95 & 69.92 & 46.19 & 62.71\\
\hline
{\bfseries Validation Accuracy (\%)} & 80.30 & {\bfseries 80.30} & 80.85 & 79.65 & 80.09 & 80.30 & 80.41\\
\hline
\end{tabular}
\end{center}
\end{table}

\subsection{Multi-Class Classification}
After binary classification, we experimented with the classification of all four labels separately. Using a learning rate of 0.0003, baseline performance of all eight models was obtained. The highest validation accuracy of 62.15\% was achieved using the “ALL” model (see Table.~\ref{tab6}). Therefore, we decided to use the “ALL” model for our multi-class experiments as well.

\begin{table}
\caption{Baseline validation accuracy - multiclass.}\label{tab6}
\begin{center}
\begin{tabular}{|l|r|}
\hline
{\bfseries Model} & {\bfseries Validation Accuracy (\%)} \\
\hline
DenseNet-121 & 56.74\\
\hline
{\bfseries ALL} & {\bfseries 62.15}\\
\hline
CheXpert & 56.08\\
\hline
MIMIC\_CH & 57.88\\
\hline
MIMIC\_NB & 57.42\\
\hline
NIH & 60.17\\
\hline
PadChest & 61.49\\
\hline
RSNA & 27.86\\
\hline
\end{tabular}
\end{center}
\end{table}

Batch size of 32 achieved the highest validation accuracy of 62.39\% (see Table.~\ref{tab7}). Hence, we decided to use a batch size of 32 for any following experiments.

\begin{table}
\caption{Batch size vs validation accuracy - multiclass.}\label{tab7}
\begin{center}
\begin{tabular}{|l|r|}
\hline
{\bfseries Batch Size} & {\bfseries Validation Accuracy (\%)}\\
\hline
8 & 62.33\\
\hline
16 & 62.09\\
\hline
{\bfseries 32} & {\bfseries 62.39}\\
\hline
64 & 62.33\\
\hline
128 & 62.15\\
\hline
\end{tabular}
\end{center}
\end{table}

Instead of assuming 0.0003 again, we used Optuna to search for the optimal learning rate with “ALL” model and batch size of 32. The highest validation accuracy of 63.18\% was reached with a learning rate of 0.0005913 (see Table.~\ref{tab8}). We estimate that the optimal learning rate is between 0.0005 and 0.0006.

\begin{table}
\caption{Learning rate vs validation accuracy - multiclass.}\label{tab8}
\begin{center}
\begin{tabular}{|l|r|}
\hline
{\bfseries Learning Rate (e-4)} & {\bfseries Validation Accuracy (\%)}\\
\hline
1.737 & 62.39\\
\hline
1.921 & 61.79\\
\hline
2.916 & 62.82\\
\hline
5.153 & 62.15\\
\hline
{\bfseries 5.913} & {\bfseries 63.18}\\
\hline
6.765 & 62.39\\
\hline
8.684 & 60.71\\
\hline
\end{tabular}
\end{center}
\end{table}

Using “ALL” model, batch size of 32, and learning rate of 0.0005913, we noticed a slight improvement using dropout. The best model achieved a validation accuracy of 63.6\% using 13.94\% dropout rate (see Table.~\ref{tab9}). We estimate that the best learning rate is between 13\% and 14\%.

\begin{table}
\caption{Dropout rate vs validation accuracy - multiclass.}\label{tab9}
\begin{center}
\begin{tabular}{|l|r|}
\hline
{\bfseries Dropout Rate (\%)} & {\bfseries Validation Accuracy (\%)}\\
\hline
3.068 &  63.30\\
\hline
4.746 & 8.516\\
\hline
8.516 & 63.24\\
\hline
{\bfseries 13.94} & {\bfseries 63.30}\\
\hline
15.82 & 63.18\\
\hline
18.59 & 63.12\\
\hline
\end{tabular}
\end{center}
\end{table}

With “ALL” model, batch size of 32, learning rate of 0.0005913, and dropout rate of 13.94, we started experimenting with augmentations. The best performing model achieved a validation accuracy of 64.38\% using +/- 1.437\% scale, +/- 5.252\% shear, and 27.61\% translation (see Table.~\ref{tab10}).

\begin{table}
\caption{Augmentations vs validation accuracy - multiclass.}\label{tab10}
\begin{center}
\begin{tabular}{|l|l|l|l|l|l|l|l|}
\hline
{\bfseries Horizontal Flip} & On & Off & On & Off & Off & {\bfseries Off} & Off\\
\hline
{\bfseries Vertical Flip} & Off & Off & Off & Off & Off & {\bfseries Off} & Off\\
\hline
{\bfseries Rotation (+/- o)} & 14 & 14 & 13 & 0 & 14 & {\bfseries 0} & 2\\
\hline
{\bfseries Scale (+/- \%)} & 13.94 & 7.98 & 5.98 & 14.47 & 5.46 & {\bfseries 1.44} & 2.87\\
\hline
{\bfseries Shear (+/- \%)} & 3.16 & 9.56 & 8.07 & 7.24 & 6.99 & {\bfseries 5.25} & 1.46\\
\hline
{\bfseries Translation (+/- \%)} & 7.36 & 18.48 & 25.08 & 25.70 & 5.17 & {\bfseries 27.61} & 28.53\\
\hline
\hline
{\bfseries Validation Accuracy (\%)} & 63.42  & 61.91 & 63.24 & 63.24 & 64.32 & {\bfseries 65.38} & 62.52\\
\hline
\end{tabular}
\end{center}
\end{table}

\begin{figure}
\includegraphics[width=\textwidth]{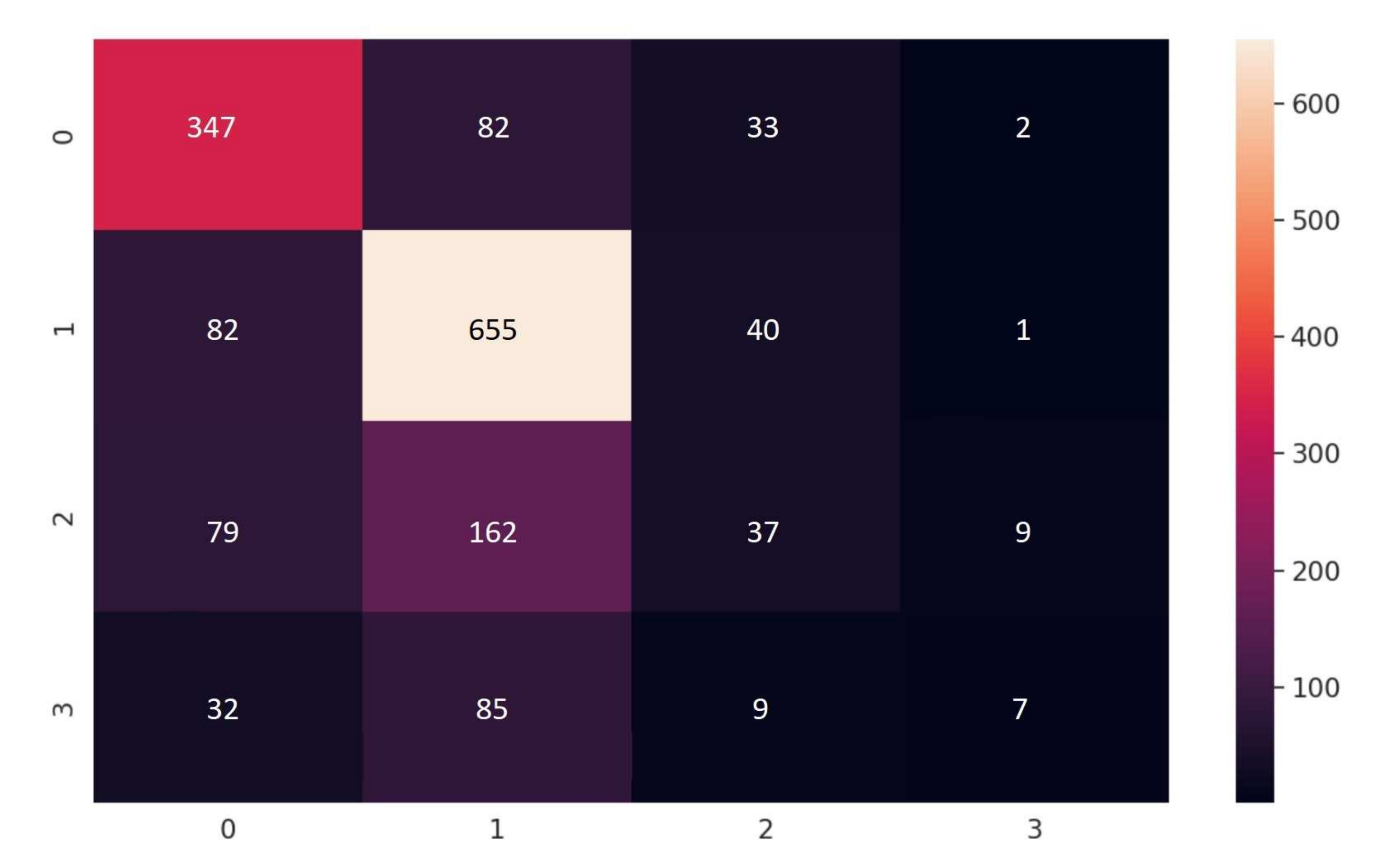}
\caption{Confusion matrix – multiclass.} \label{fig1}
\end{figure}

\begin{figure}
\includegraphics[width=\textwidth]{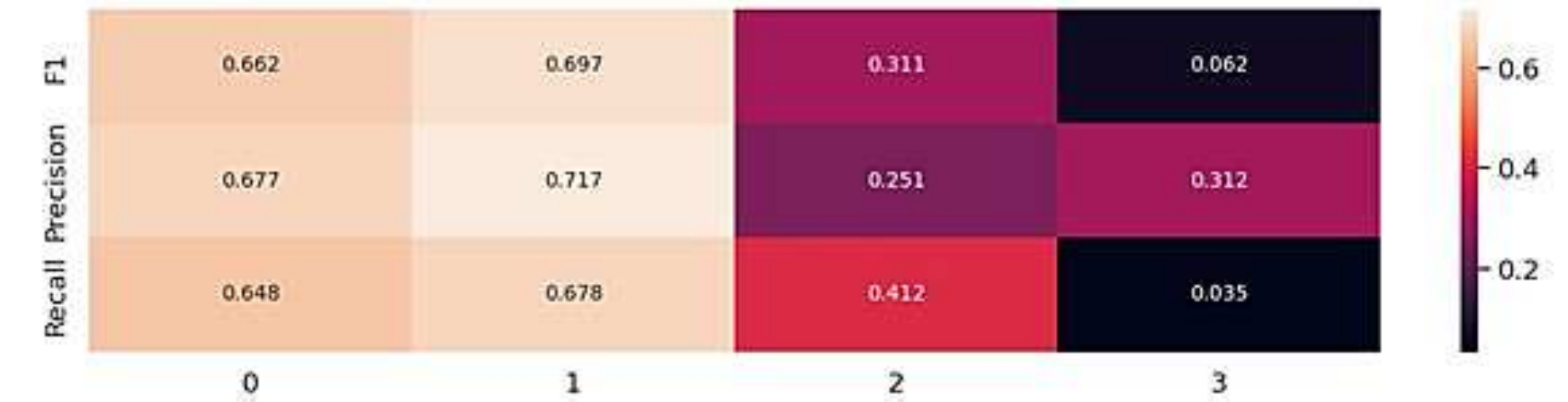}
\caption{F1 heatmap - multiclass.} \label{fig2}
\end{figure}

As shown above, the accuracy of our multi-class models hardly exceeded the 60\% mark before overfitting the data. After inspecting our confusion matrices (see Fig.~\ref{fig1}) and F1 scores (see Fig.~\ref{fig2}), we discovered that the models are simply unable to recognize the third (‘Indeterminate Appearance’) and forth (‘Atypical Appearance’) classes, especially the latter. This might be a direct cause of having a smaller amount of data for them compared to the other two.

\section{Conclusion}
According to a study~\cite{ref_paper5}, up to 29\% of negative PCR tests can become positive when repeated. With careful tuning of augmentations, hyperparameters, and model, we are able to produce a model with 83\% accuracy that can be used as a quick way to detect COVID-19 early and take precautionary measure even without a positive PCR result. However, for multi-class classification, even with further hyperparameter optimization, we don’t believe it is possible to exceed an accuracy of 65-66\%. In other words, we can classify the presence of COVID-19 pneumonia from chest x-rays but the severity of the infection.

Augmentations are widely accepted as a method to improve accuracy, especially when there is lack or imbalance of data~\cite{ref_paper3}. However, in the case of our binary experiments, this concept did not hold true. Adding dropout was also detrimental to the performance of our binary models. In fact, it caused the training accuracy to peak then decrease, preventing the model from overfitting. 

After examining the final results of the competition, we discovered that the highest score achieved was 63.5\%\cite{ref_url16}. DICE score was used because the challenge asked for bounding boxes and not just classification. With that score in mind, we believe that there is a limitation within the dataset itself regardless of model, learning rate, dropout, batch size, and/or augmentation selection. For future work, we would still like to experiment with other models, optimizers, losses, as well as GAN-based data augmentations\cite{ref_paper8} to improve our multi-class model.

\bibliographystyle{splncs04}
\bibliography{mybibliography}

\appendix
\section{Appendix}

\begin{figure}
\includegraphics[width=\textwidth]{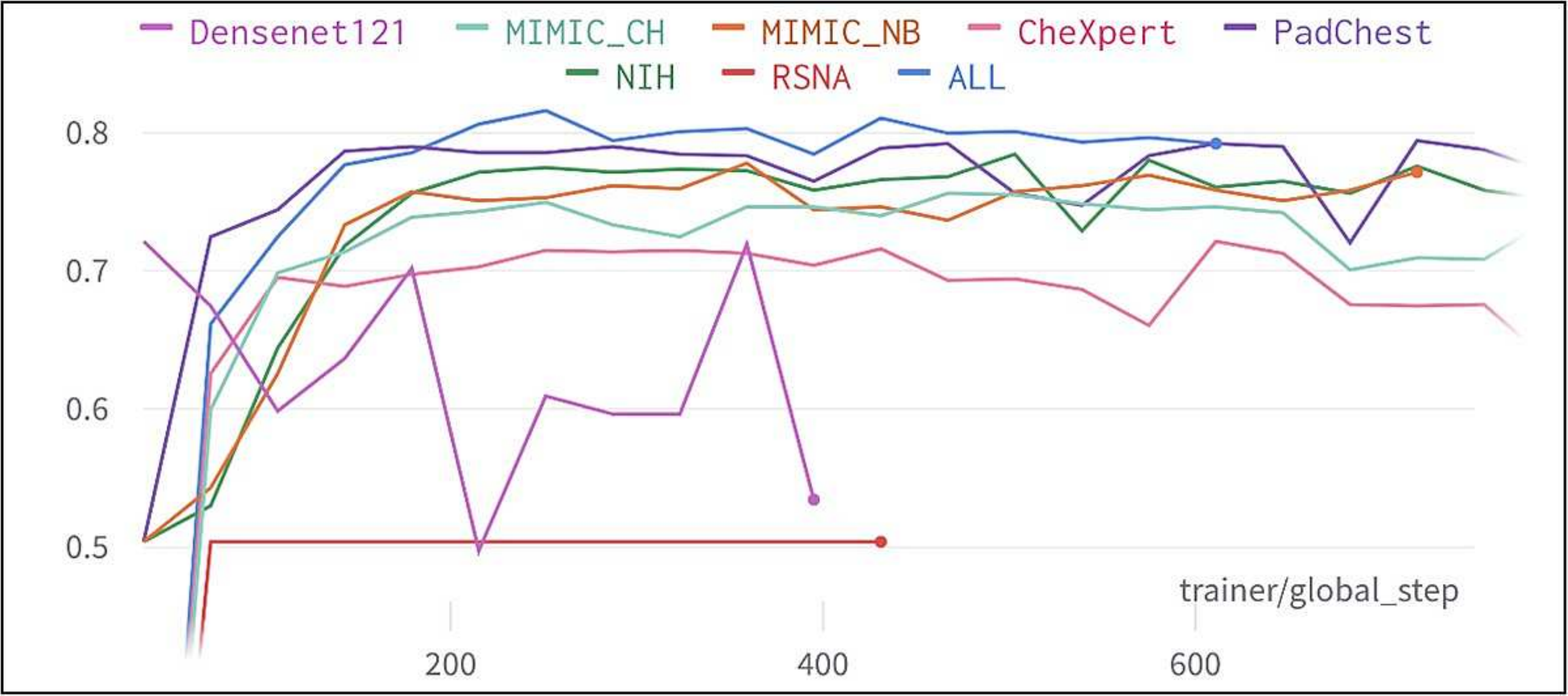}
\caption{Model vs validation accuracy – binary.} \label{fig3}
\end{figure}

\begin{figure}
\includegraphics[width=\textwidth]{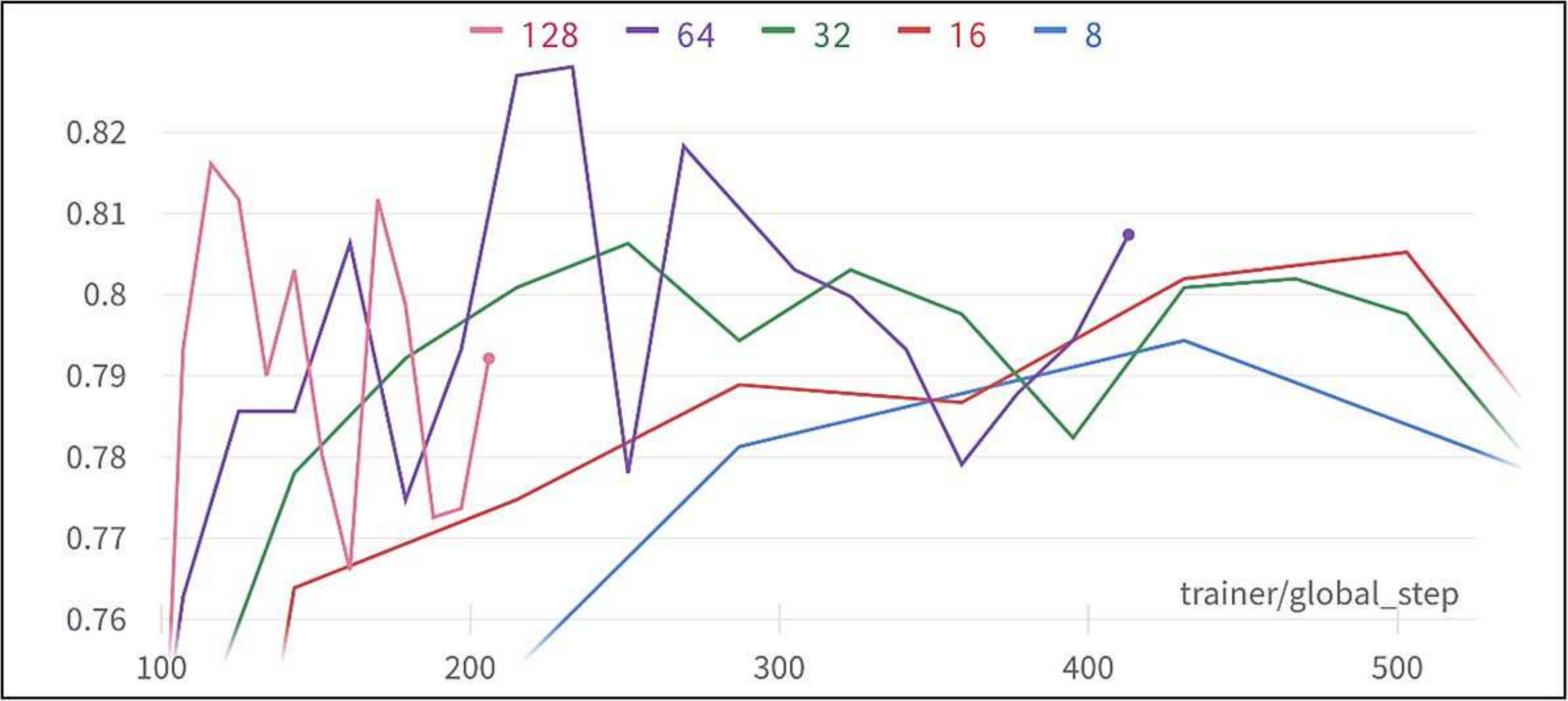}
\caption{Batch size vs validation accuracy - binary.} \label{fig4}
\end{figure}

\begin{figure}
\includegraphics[width=\textwidth]{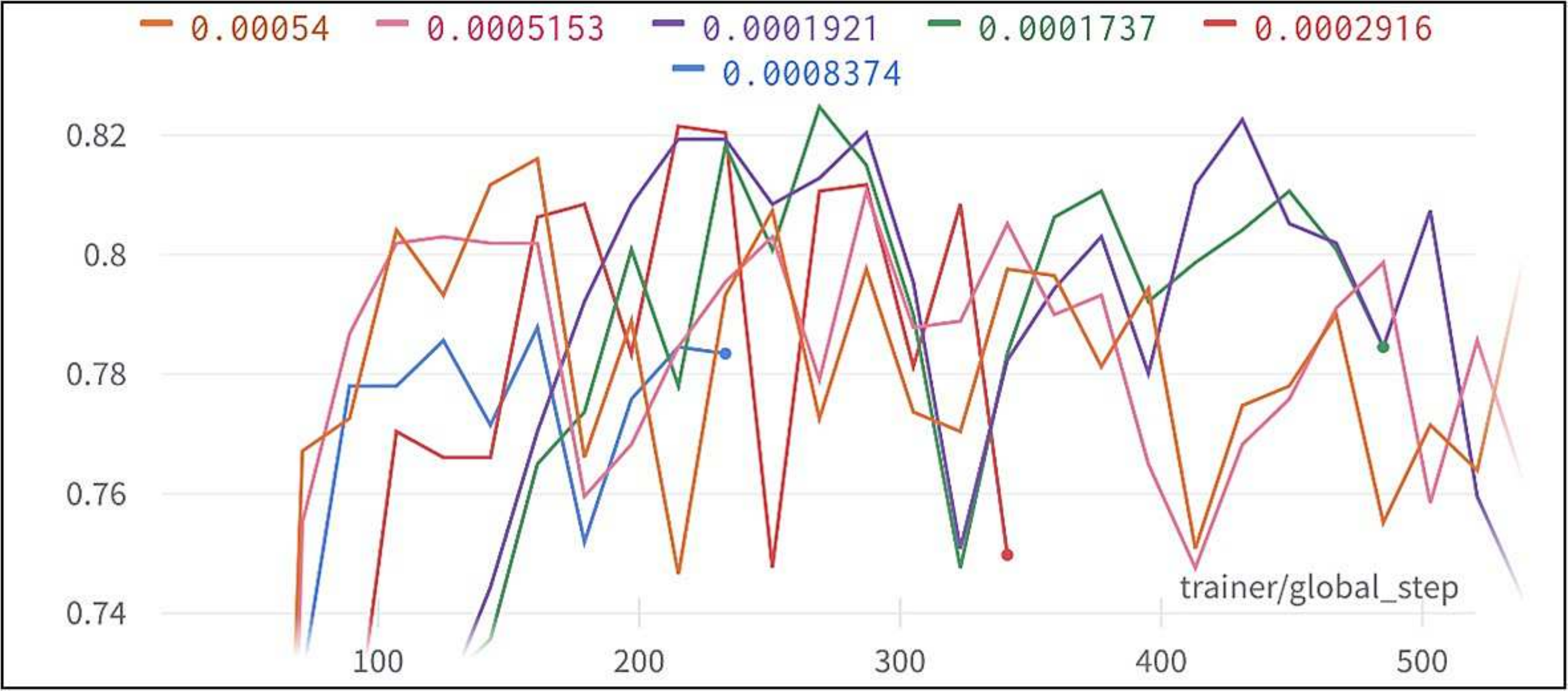}
\caption{Learning rate vs validation accuracy – binary.} \label{fig5}
\end{figure}

\begin{figure}
\includegraphics[width=\textwidth]{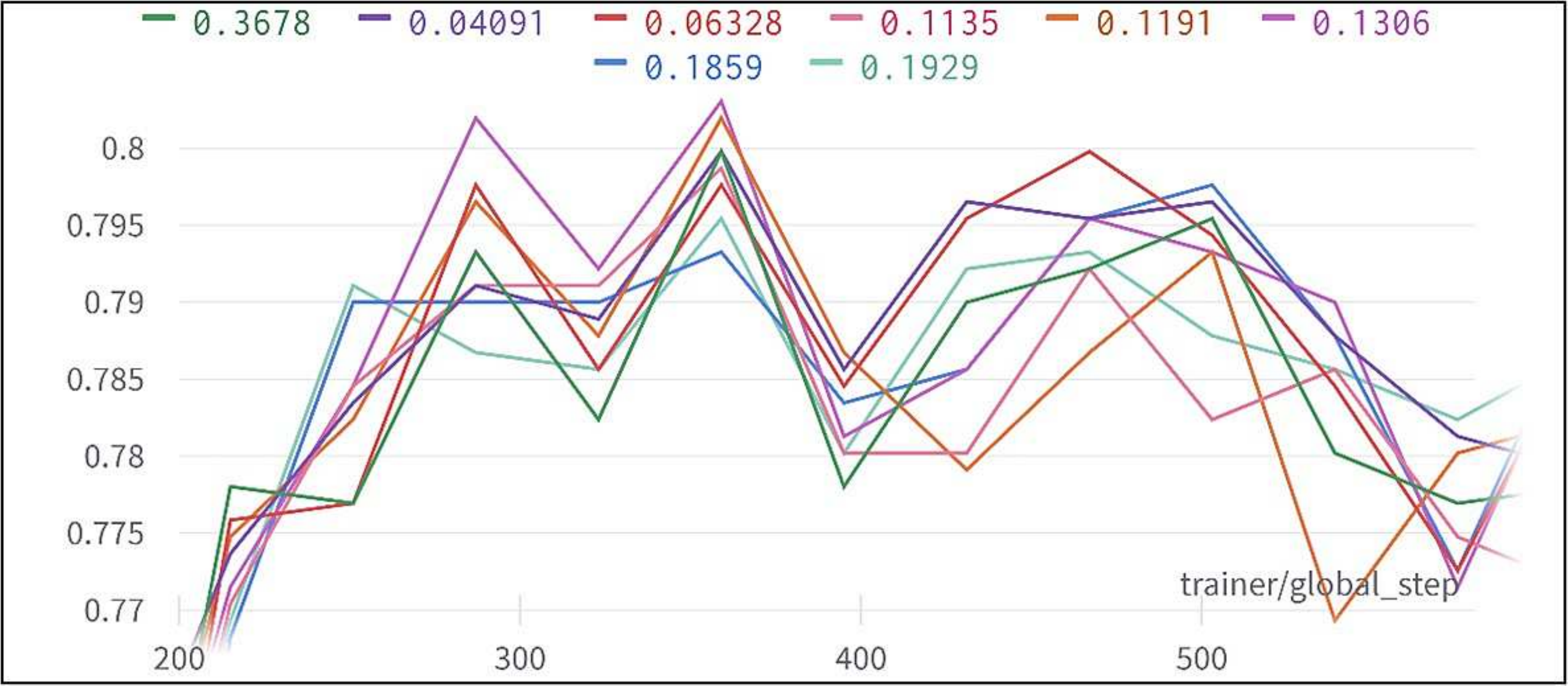}
\caption{Dropout rate vs validation accuracy - binary.} \label{fig6}
\end{figure}

\begin{figure}
\includegraphics[width=\textwidth]{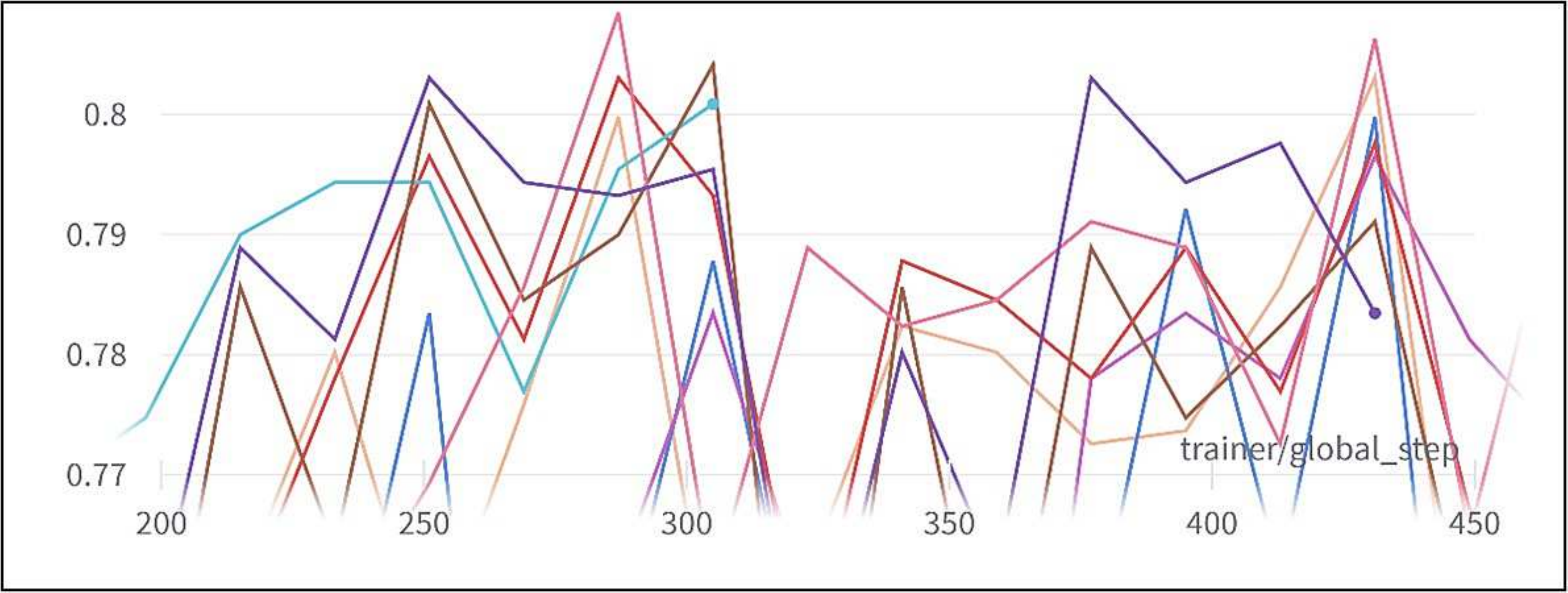}
\caption{Top 8 augmentations vs validation accuracy - binary.} \label{fig7}
\end{figure}

\begin{figure}
\includegraphics[width=\textwidth]{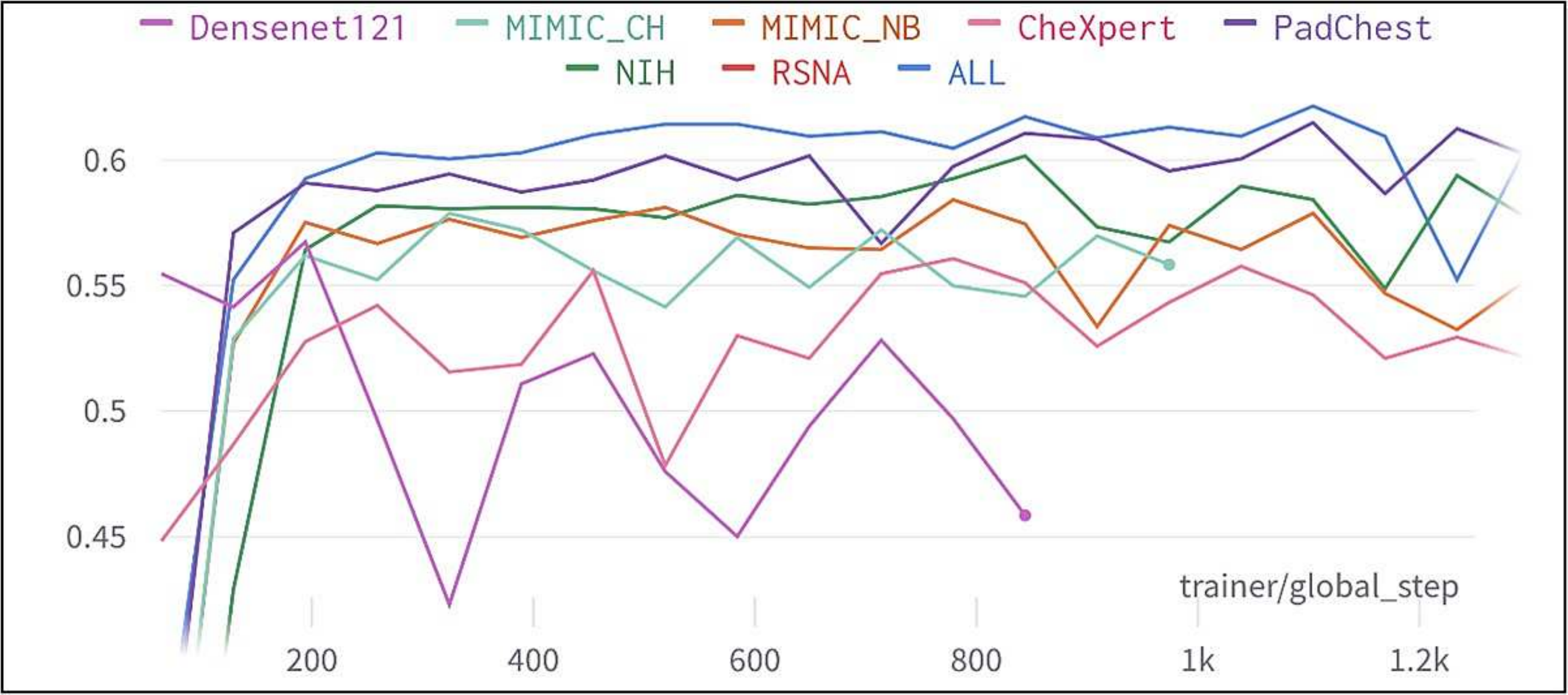}
\caption{Model vs validation accuracy - multiclass.} \label{fig8}
\end{figure}

\begin{figure}
\includegraphics[width=\textwidth]{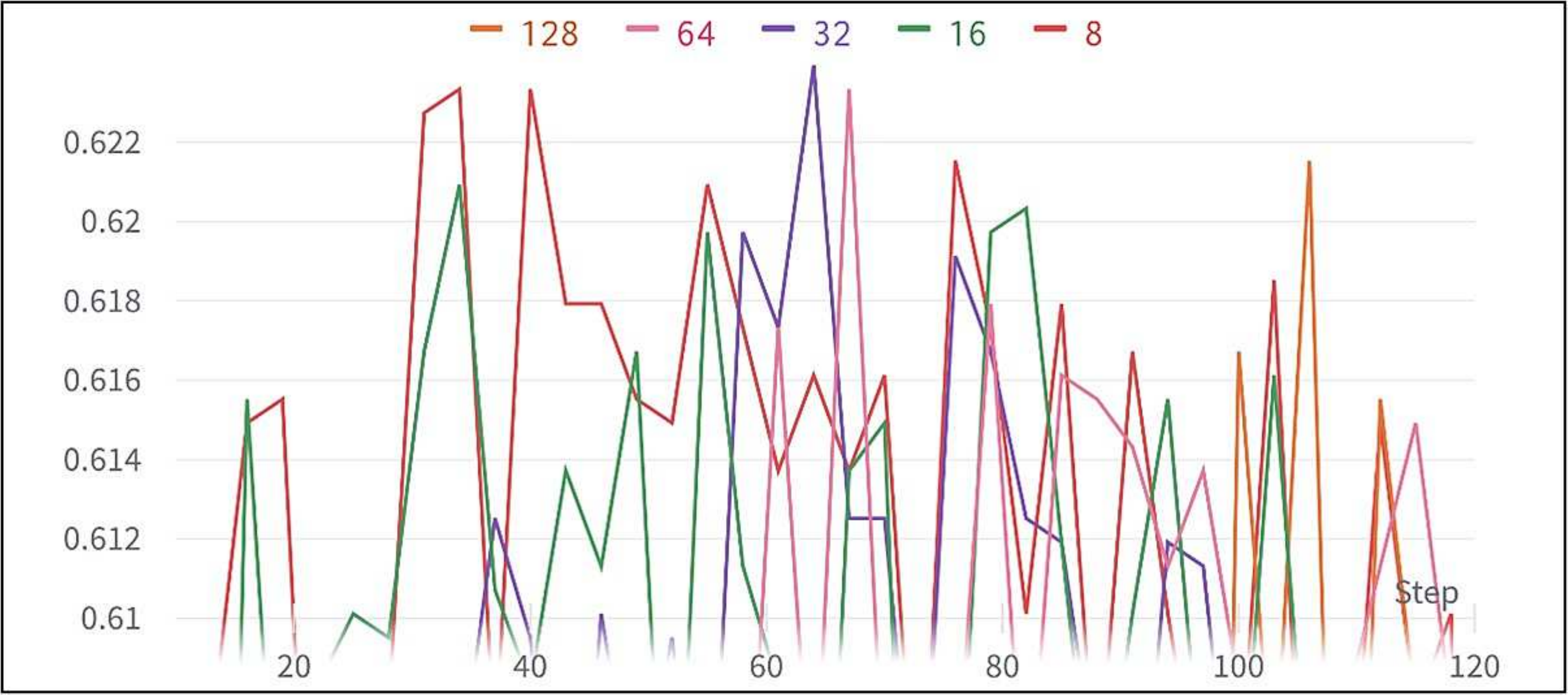}
\caption{Batch size vs validation accuracy - multiclass.} \label{fig9}
\end{figure}

\begin{figure}
\includegraphics[width=\textwidth]{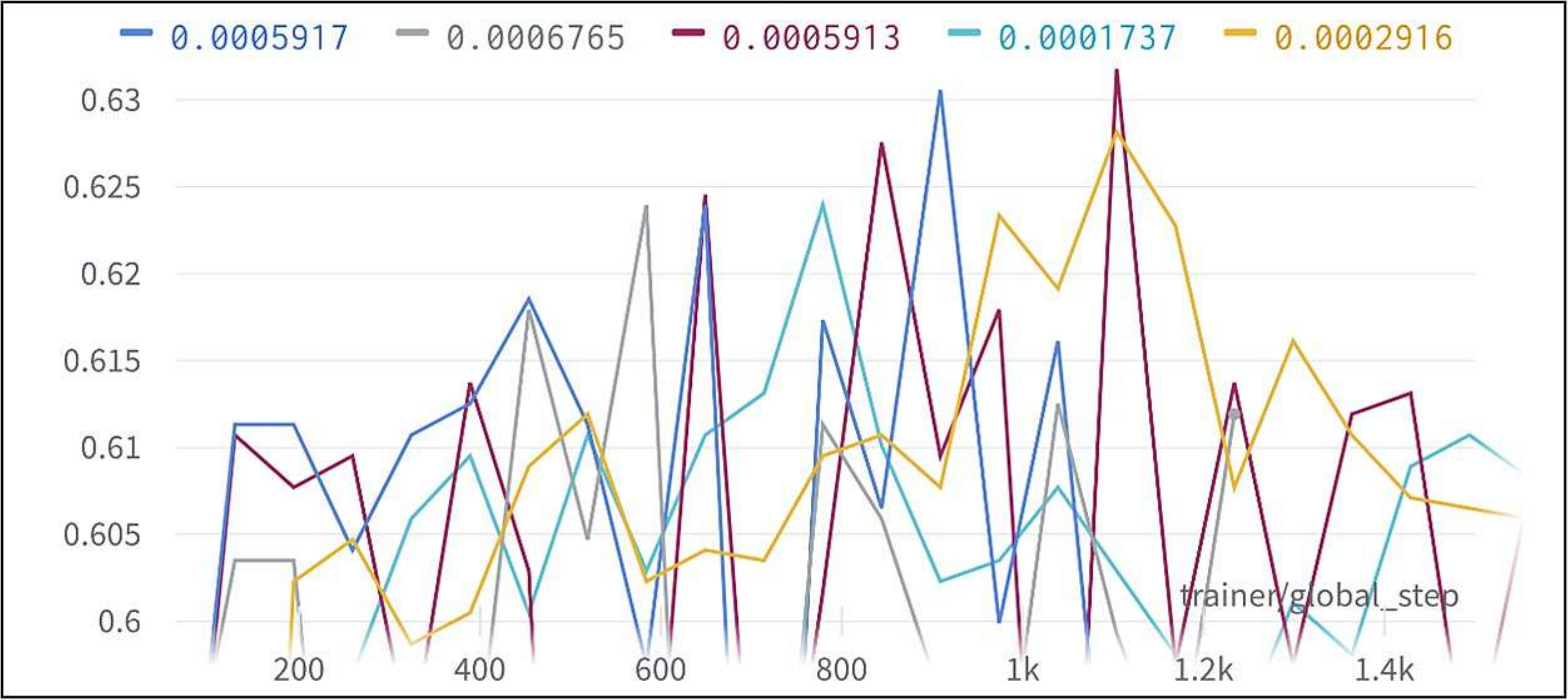}
\caption{Learning rate vs validation accuracy - multiclass.} \label{fig10}
\end{figure}

\begin{figure}
\includegraphics[width=\textwidth]{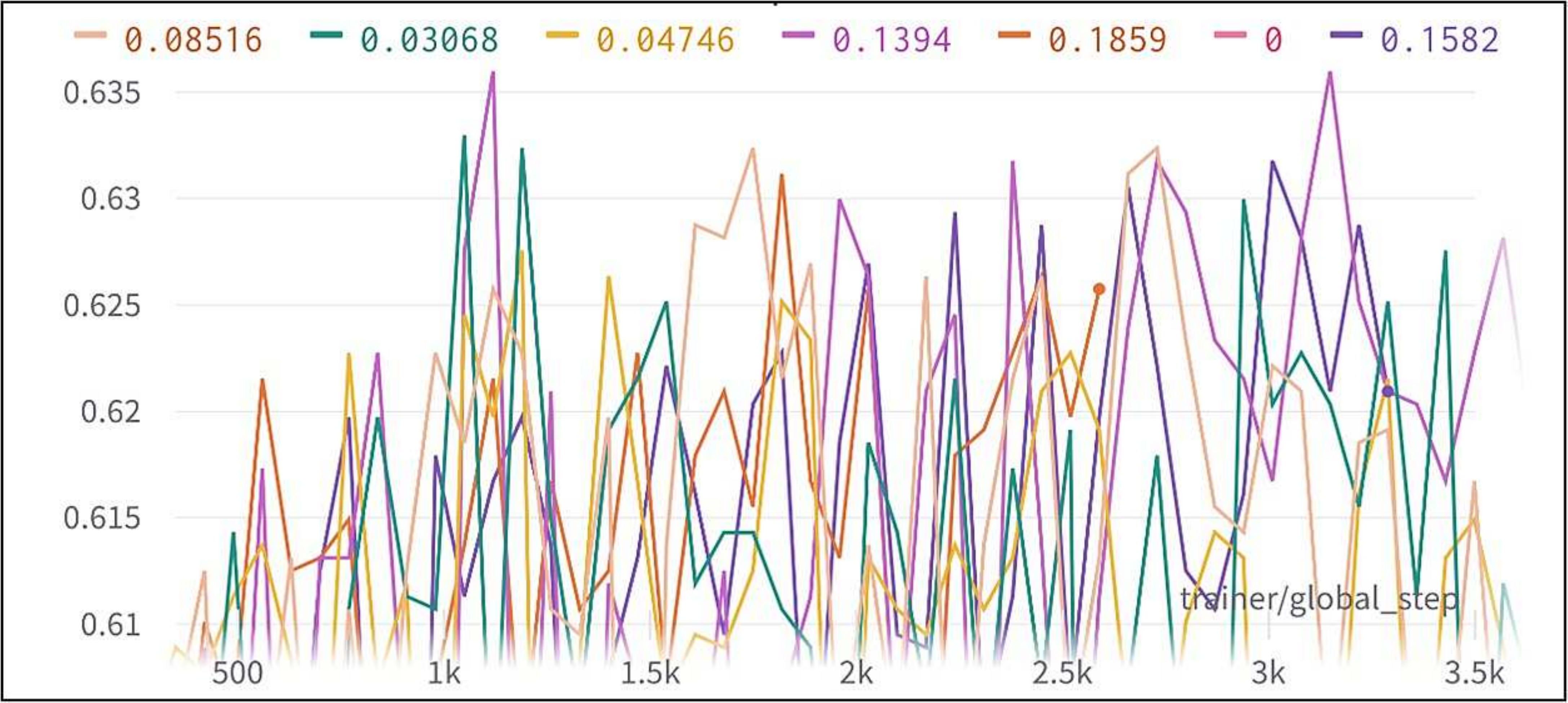}
\caption{Dropout rate vs validation accuracy – multiclass.} \label{fig11}
\end{figure}

\begin{figure}
\includegraphics[width=\textwidth]{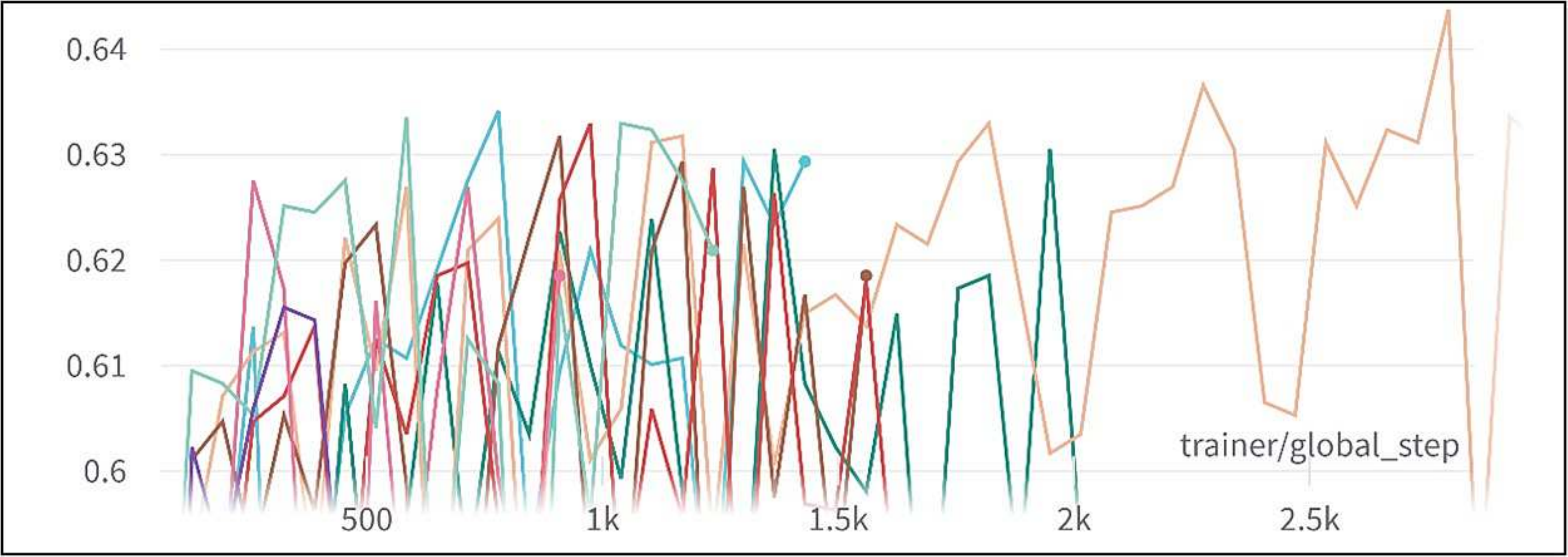}
\caption{Top 8 augmentations vs validation accuracy – multiclass.} \label{fig12}
\end{figure}

\begin{figure}
\includegraphics[width=\textwidth]{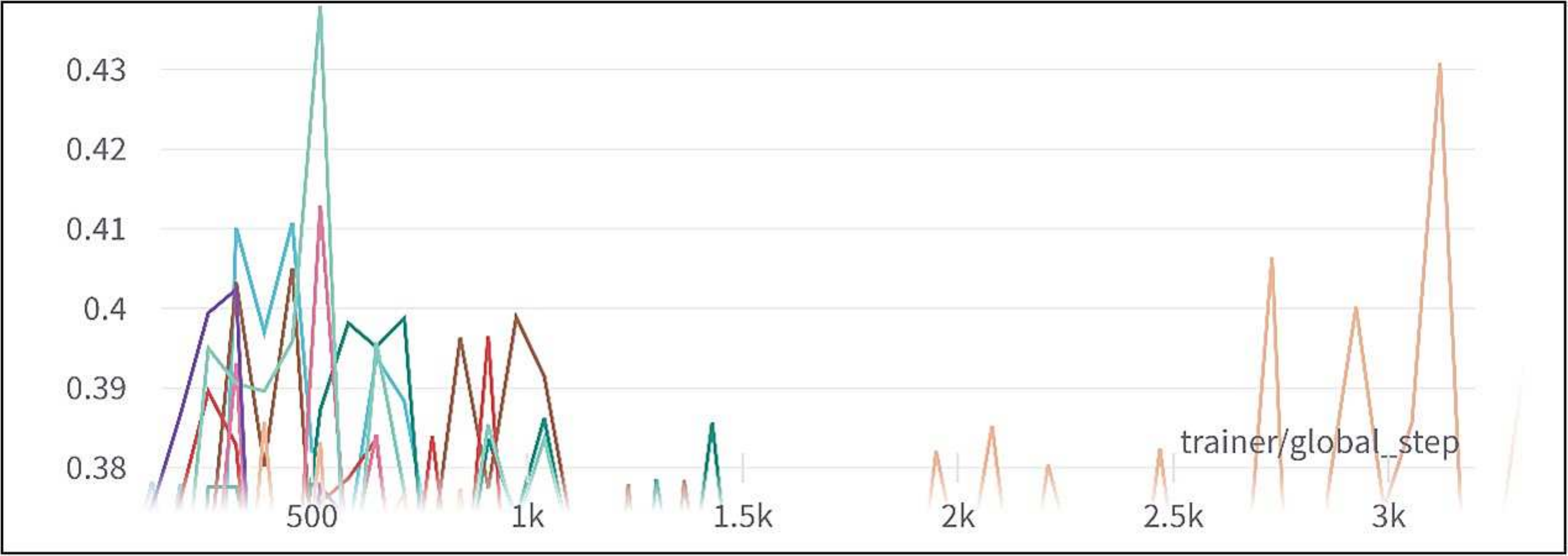}
\caption{Top 8 augmentations vs f1 macro score - multiclass.} \label{fig13}
\end{figure}

\end{document}